\documentclass[journal=NanoLetters,manuscript=article]{achemso}
\pdfoutput=1
\setkeys{acs}{keywords = true}
\setkeys{acs}{etalmode=truncate,maxauthors=10}
\usepackage[version=3]{mhchem} 
\usepackage[T1]{fontenc}       
\usepackage{multirow}

\author{Changhua Bao}
\affiliation
{State Key Laboratory of Low Dimensional Quantum Physics and Department of Physics, Tsinghua University, Beijing 100084, P.R. China}
\author{Wei Yao}
\affiliation
{State Key Laboratory of Low Dimensional Quantum Physics and Department of Physics, Tsinghua University, Beijing 100084, P.R. China}
\author{Eryin Wang}
\affiliation
{State Key Laboratory of Low Dimensional Quantum Physics and Department of Physics, Tsinghua University, Beijing 100084, P.R. China}
\author{Chaoyu Chen}
\affiliation
{Synchrotron SOLEIL, L'Orme des Merisiers, Saint Aubin-BP 48, 91192 Gif sur Yvette Cedex, France}
\author{Jos\'{e} Avila}
\affiliation
{Synchrotron SOLEIL, L'Orme des Merisiers, Saint Aubin-BP 48, 91192 Gif sur Yvette Cedex, France}
\author{Maria C. Asensio}
\affiliation
{Synchrotron SOLEIL, L'Orme des Merisiers, Saint Aubin-BP 48, 91192 Gif sur Yvette Cedex, France}
\author{Shuyun Zhou}
\affiliation
{State Key Laboratory of Low Dimensional Quantum Physics and Department of Physics, Tsinghua University, Beijing 100084, P.R. China}
\alsoaffiliation
{Collaborative Innovation Center of Quantum Matter, Beijing 100084, P.R. China}
\email{syzhou@mail.tsinghua.edu.cn}
\phone{+86 010 62797928}
\title[An \textsf{achemso} demo]
  {Stacking-dependent electronic structure of trilayer graphene resolved by nanospot angle-resolved photoemission spectroscopy \footnote{The authors declare no competing financial interest.}}


\begin{document}


\begin{tocentry}
\centering
  \includegraphics[]{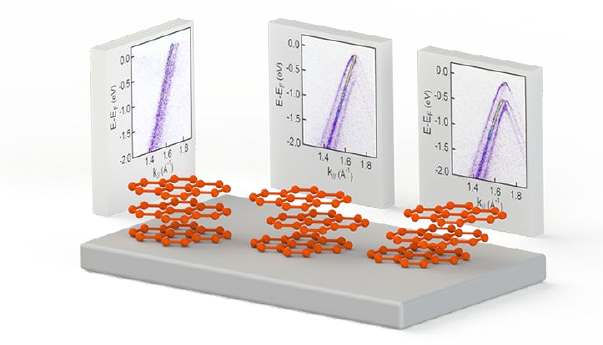}

\end{tocentry}
\begin{abstract}
  The crystallographic stacking order in multilayer graphene plays an important role in determining its electronic structure.  In trilayer graphene, rhombohedral stacking (ABC) is particularly intriguing, exhibiting a flat band with an electric-field tunable band gap. Such electronic structure is distinct from simple hexagonal stacking (AAA) or typical Bernal stacking (ABA), and is promising for nanoscale electronics, optoelectronics applications. So far clean experimental electronic spectra on the first two stackings are missing because the samples are usually too small in size ($\mu$m or nm scale) to be resolved by conventional angle-resolved photoemission spectroscopy (ARPES).  Here by using ARPES with nanospot beam size (NanoARPES), we provide direct experimental evidence for the coexistence of three different stackings of trilayer graphene and reveal their distinctive electronic structures directly. By fitting the experimental data, we provide important experimental band parameters for describing the electronic structure of trilayer graphene with different stackings.

\end{abstract}


Graphene, a monolayer of carbon atoms arranged in a honeycomb lattice, has been extensively investigated in the last decade as a two-dimensional material with intriguing  properties originated from its linearly dispersing Dirac cone at the K point \cite{Graphene2007}.  When stacking monolayer graphene to form multilayer graphene, the interlayer interaction can lead to dramatic changes of the Dirac cone, depending on how the graphene layers are stacked. The simplest example is bilayer graphene, which has two different stacking sequences, AB (Bernal) and AA stackings. In contrast to bilayer graphene with AA stacking  which shows two linearly dispersing Dirac cones displaced from the K point, bilayer Bernal graphene shows parabolic dispersions. A band gap can be induced by a perpendicular electric field, making it potentially useful for applications in electronics and photonics  \cite{TheoryBand2007,bilayerOostinga2008,bilayerOhta2006}.

\begin{figure*}[htbp]
  \centering
  \includegraphics[]{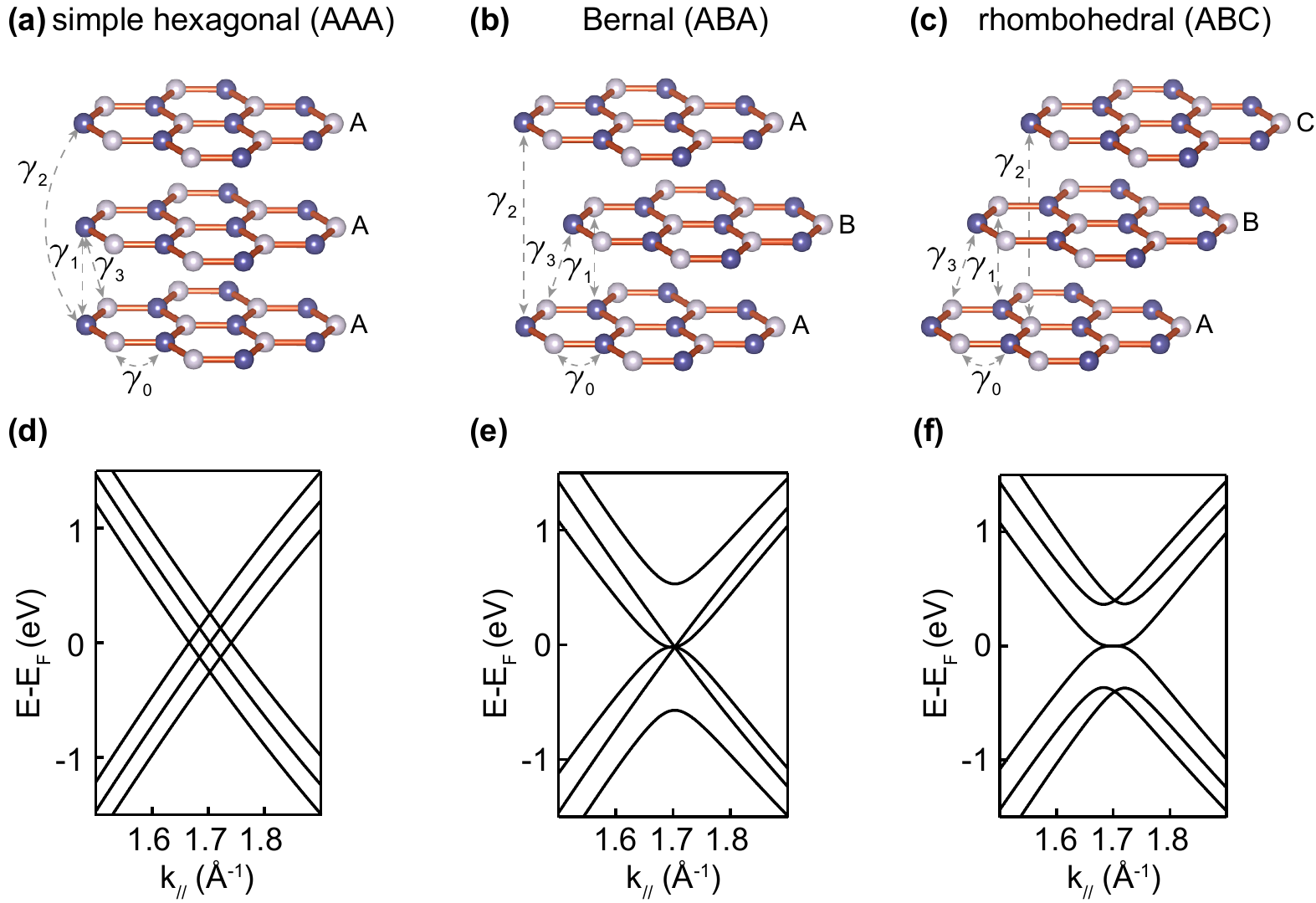}
  \caption{Three different stackings for trilayer graphene and the corresponding calculated electronic structures. (a)-(c) Schematic drawings of simple hexagonal (AAA), Bernal (ABA) and rhombohedral (ABC) stackings. The main hopping parameters are labeled.  (d)-(f)Theoretical $\pi$ band dispersions of AAA, ABA and ABC stackings along the $\Gamma$-K-M direction using tight binding model with hopping terms $\gamma_0$, $\gamma_1$ and $\gamma_3$.
}\label{fig1}
\end{figure*}

In trilayer graphene, the different stacking sequences provide an even richer playground for electronic band structure engineering \cite{TheoryBand2007}. There are three stacking sequences, simple hexagonal (AAA), Bernal (ABA) and rhombohedral (ABC) stackings as schematically shown in Fig.~\ref{fig1}(a)-(c). The different stacking leads to different vibrational and electronic properties. The Raman and infrared active modes, electron-phonon coupling are quite different for ABA and ABC stacking sequences which has also been used as a reliable and efficient method to determine the stacking sequences \cite{Lui2011Imaging,Raman1,Raman2,Raman3,Raman4}. Moreover, they have different response under applied electric field. AAA stacking has the highest symmetry, and its electronic band structure consists of three equally displaced Dirac cones as shown in Fig. \ref{fig1}(d).  Applying an electric field can increase the separation between the Dirac cones, however the Dirac cones remain gapless \cite{Lu2007,Chang2008}. The most common ABA stacking has mirror symmetry and lacks inversion symmetry. Its band structure shows effectively the superposition of a linear Dirac cone from monolayer graphene and two quadratic dispersions from AB stacking bilayer graphene (see Fig.~\ref{fig1}(e)) \cite{TheoryBand2008PRB,Latil2006,TheoryBand2007,Guinea2007,TheoryBand2008,TheoryBand2010}.  Applying an electric field will induce a gap only for the linear dispersing band, while the parabolic bands will still remain gapless \cite{Craciun2009,koshino2009gate,TheoryBand2010}. Because of the absence of a band gap even under an applied electric field, both of these two graphene stackings are not very useful for electronic devices.

Rhombohedral stacking (ABC) graphene has inversion symmetry, but lacks mirror symmetry. The  crystallographic symmetry leads to two flat bands at the Fermi level with cubic dispersion (Fig.~\ref{fig1}(f))  \cite{Latil2006,TheoryBand2007,Guinea2007,Lu2007,Chang2008,TheoryBand2008PRB,TheoryBand2008,
TheoryBand2010,TheoryBandABC2009,TheoryBandABC2010}.  The double degeneracy in rhombohedral stacking graphene can be lifted by applying different potentials to the top and bottom graphene layers  \cite{Lu2007,Avetisyan2009,TheoryBandABC2010,TheoryBand2010,gap2010}.  Experimentally, evidence of existence of the tunable gap has been inferred by infrared conductivity \cite{gap2011}, electrical and magnetic transport measurements \cite{Bao2011}.
Another important property of rhombohedral trilayer graphene is that the band at the K point near the Fermi level has very small velocity, and in the limit of many layers will become a flat band  \cite{flatband2013,flatband2011,flatband2015}. The high density of states from the flat band provides new opportunities for realizing many exotic properties, e.g. flat band high temperature superconductivity \cite{superScheike2012,flatband2011,
flatband2013}, various magnetic orders  \cite{Otani2010a} etc.  Under an applied magnetic field, a Lifshitz transition induced by trigonal warping in rhombohedral trilayer graphene has been reported  \cite{Bao2011,TheoryBandABC2009}. It has also been predicted that ferromagnetic spin polarization can exist on the (0001) surfaces of rhombohedral graphite \cite{Otani2010}, and a thin film of rhombohedral graphene can  undergo a magnetic phase transition from antiferromagnetic state to the ferromagnetic state under a perpendicular electric field  \cite{Otani2010a}. In brief, rhombohedral trilayer graphene provides a platform to investigate very rich physics and promising applications in electronics, optoelectronics and so on.

Energetically, rhombohedral stacking is less stable than Bernal stacking \cite{TheoryBand2007}. Although rhombohedral graphene has been identified in exfoliated graphene \cite{Lui2011Imaging} and multilayer graphene grown on 3C-SiC(111) or 6H-SiC(0001) substrate \cite{flatband2015,ARPES2013,ARPES2007}, rhombohedral graphene is usually mixed with the dominant Bernal graphene, and the small grain size makes it challenging to obtain clean electronic band dispersions \cite{ARPES2007,ARPES2013}. The epitaxial graphene sample was grown by annealing the Pt(111) substrate in ultrahigh vacuum at elevated temperatures up to 1600 $^{\circ}$C using electron beam bombardment \cite{Yao2015,Sutter, GaoHJ}. Using nanospot angle-resolved photoemission spectroscopy (NanoARPES), we are able to record electronic nanoimages which reveal different density of states close to the Fermi level related with distinctive regions of different stacking sequences. We resolved the ABA and ABC stacking orders completely and obtained very clean and high quality electronic bands compared to previous reports \cite{ARPES2007,ARPES2013}. We also obtained the band structure of AAA stacked trilayer graphene for the first time. The hopping parameters are extracted using tight-binding fitting. Our work reveals the three dissimilar dispersions relation for the three trilayer graphene stacking sequences, and provides important experimental band parameters for describing the electronic structure of trilayer graphene.

\begin{figure*}[htbp]
  \centering
  \includegraphics[]{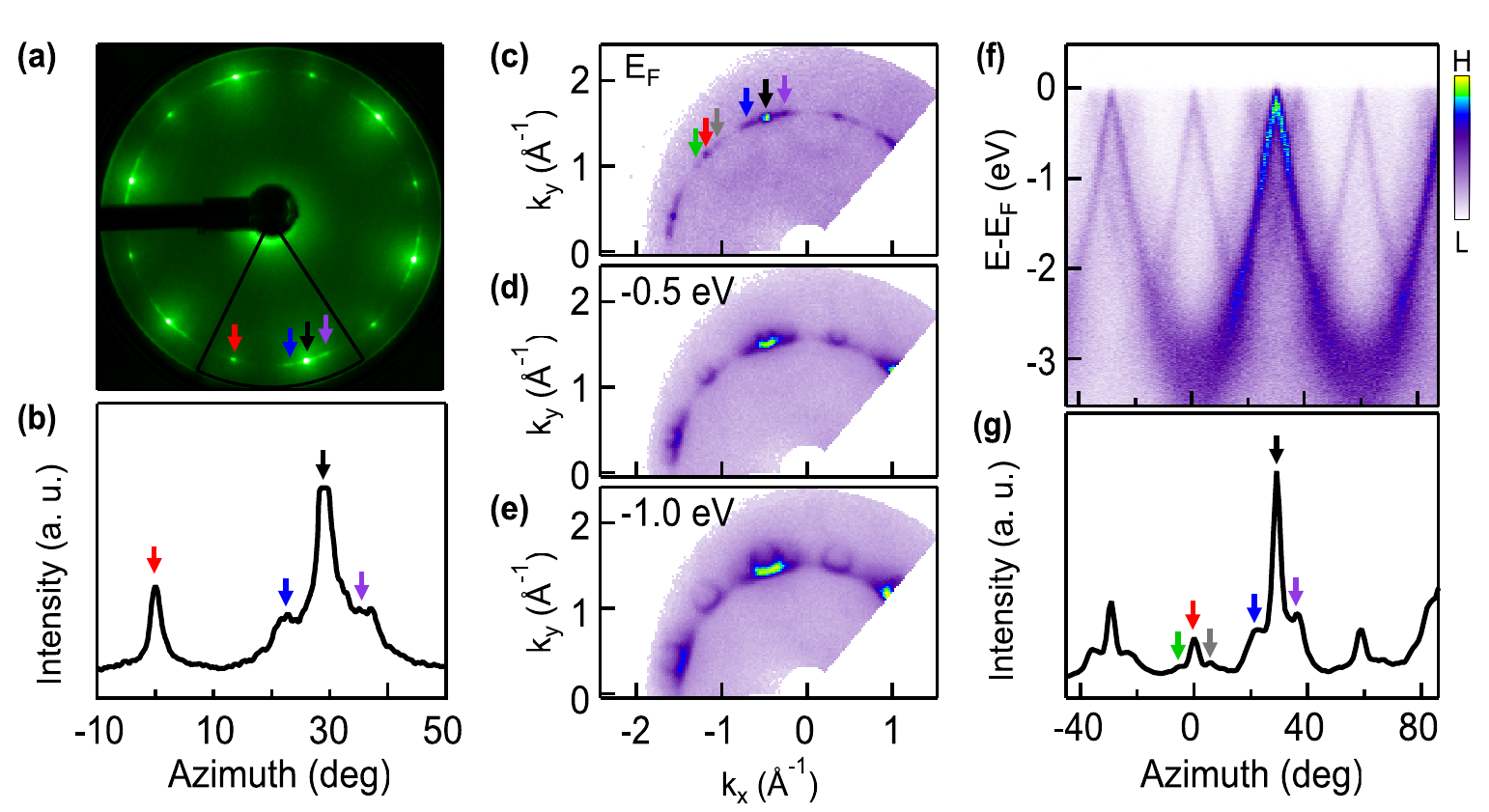}
  \caption{Different orientations of graphene on (111) surface of platinum. (a) LEED pattern of multilayer graphene on platinum. (b) The azimuth dependence of LEED intensity. (c)-(d) Constant energy maps at Fermi energy, 0.5 eV and 1 eV below Fermi energy respectively. (f) A circular cut through the K-points. (g) The azimuth dependence of intensity along the circle through the K-points at Fermi energy.}\label{fig2}
\end{figure*}

Figure \ref{fig2} shows the characterization of the orientation of multilayer graphene relative to the substrate using low-energy electron diffraction (LEED) and ARPES. Figure \ref{fig2}(a) shows the LEED pattern. The most obvious features are indicated by red and black arrows, which have  0$^\circ$ and 30$^\circ$ azimuthal angles relative to the platinum substrate. Besides, there is some arc-like shape  around these two patterns. This can be seen more clearly in the azimuthal dependence curve in Fig.~\ref{fig2}(b). In addition to 0$^\circ$ and 30$^\circ$ grains, there exist 23.4$^\circ$ and 36.6$^\circ$ grains indicated by blue and purple arrows respectively, and shoulders around the 0$^\circ$ peak in Fig.~\ref{fig2}(b). The Fermi surface map obtained by regular ARPES with beam size of $\approx$ 100 $\mu$m is shown in Fig.~\ref{fig2}(c), and their evolutions at -0.5 eV and -1.0 eV are shown in Fig.~\ref{fig2}(d) and (e) respectively which show the Dirac cone features with different orientations. The azimuth dispersion (Fig.~\ref{fig2}(f)) and the momentum distribution curve at E$_F$ (Fig.~\ref{fig2}(g)) reveal domains with orientations consistent with the LEED pattern. But more orientations can be distinguished in ARPES spectrum which are hidden in the shoulder of the LEED data and indicated by gray (6$^\circ$) and green (-6$^\circ$) arrows. Combining LEED and regular ARPES data, we reveal at least six different orientations of graphene grown on (111) surface of platinum, 0$^\circ$, 6$^\circ$, 23.4$^\circ$, 30$^\circ$, 36.6$^\circ$ and 54$^\circ$ (equivalent to -6$^\circ$). The rich orientations are resulted from the weak graphene-substrate interaction, which is also likely to lead to  different stacking sequences.

\begin{figure*}[htbp]
  \centering
  \includegraphics[]{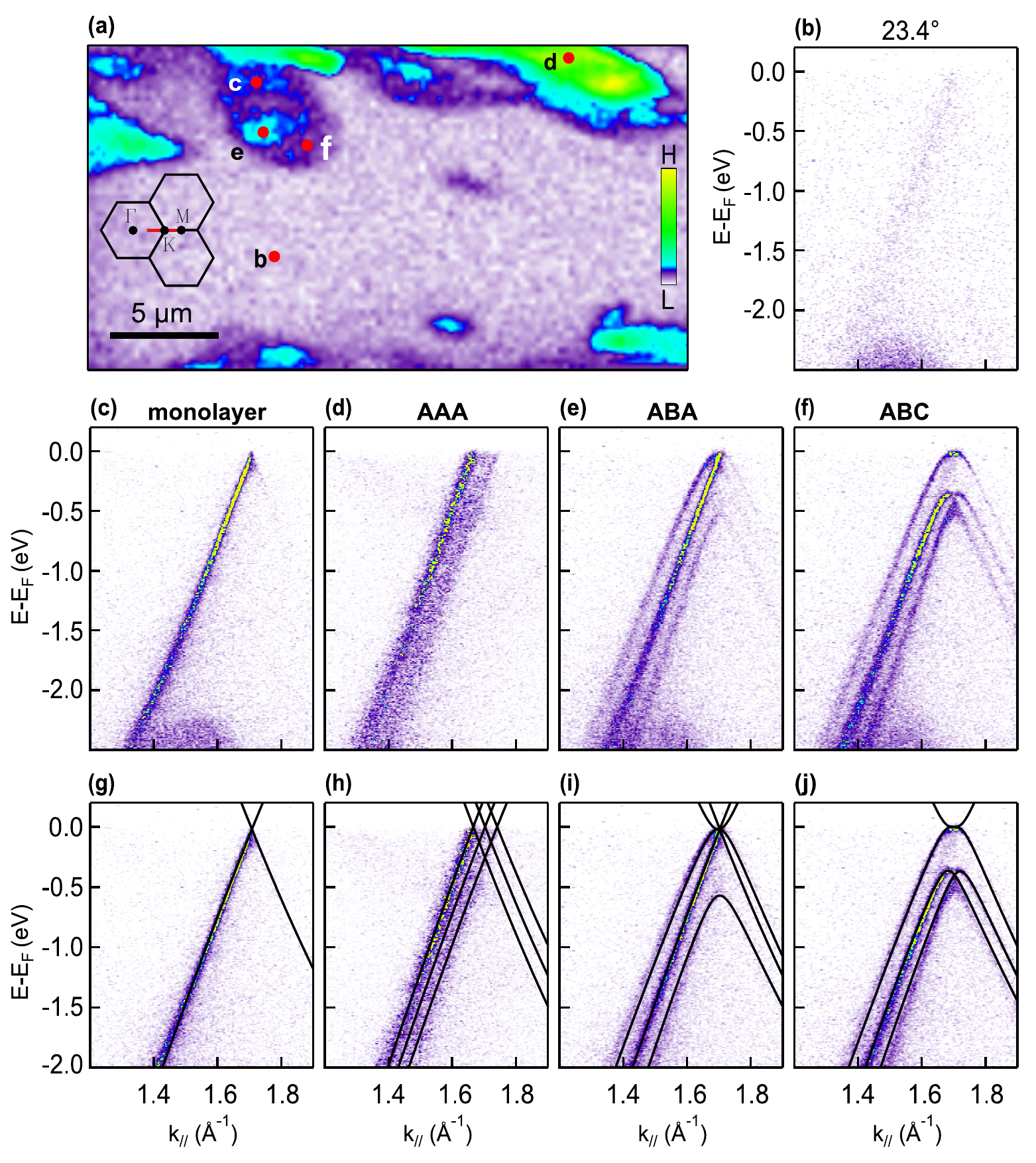}
  \caption{Spatially resolving nanometer scale regions of different stacking orders and the corresponding band structure. (a) A NanoARPES image by integrating intensity from -0.1 eV to -0.5 eV. The inset shows a schematic drawing of the Brillouin zone and the direction of measurement for 0 degree domain. (b)-(f) NanoARPES spectra at $\phi=0^\circ$ along the $\Gamma$-K-M direction and at h$\nu$=100 eV, from five different regions at the correspondingly labeled positions in (a). (g)-(j) Fit results for monolayer, simple hexagonal, Bernal and rhombohedral trilayer graphenes respectively by using theoretical tight-binding model.}\label{fig3}
\end{figure*}

To resolve the electronic structure of multilayer graphene, we perform NanoARPES measurements. Figure \ref{fig3}(a) shows the spatially resolved map integrated from -0.1 to -0.5 eV along the $\Gamma-K-M$ direction of 0$^\circ$ grain.  Figure \ref{fig3}(b)-(f) shows NanoARPES data taken from five different typical regions marked in panel (a). The large region marked by label b shows negligible intensity within -2 eV of E$_F$ and there is a peak at $\approx$ -2.2 eV. This is identified as 23.4$^\circ$ graphene.  The existence of this peak also in panels (c), (e) and (f) suggests that these graphene domains are on 23.4$^\circ$ graphene. In Fig.~\ref{fig3}(c), the linear dispersion from 0$^\circ$ monolayer graphene on 23.4$^\circ$ graphene is very clear. The asymmetrical intensity is due to matrix element effect.\cite{ARPESReview}  Dispersions of characteristic trilayer graphene are observed in regions marked by label d, e and f.
In Fig. \ref{fig3}(d), three linear dispersions are present, and they are identified to be from simple hexagonal trilayer graphene. A linear dispersion band and two quadratic dispersion bands from Bernal  trilayer graphene are shown clearly in Fig. \ref{fig3}(e). Figure \ref{fig3}(f) shows two intersected quadratic dispersion bands below E$_F$ and a flat band at E$_F$ from rhombohedral trilayer graphene.  Therefore, in the area which we studied, there are monolayer graphene, trilayer graphene with simple hexagonal, Bernal and rhombohedral stackings. Every trilayer graphene domain shows very distinct dispersion associated to its corresponding stacking.

\begin{table*}
\caption{Hopping parameters  (in eV) and Fermi velocity (in 10$^6$m/s) extracted from fitting NanoARPES spectra using tight-binding model with $\gamma_0$, $\gamma_1$ and $\gamma_3$ as compared to previous reports.}\label{fit}
\begin{tabular}{c  c c c c c c c c}
  \hline
  \hline
  stacking  & $v_F$ & $\gamma_0$ & $\gamma_1$ & $\gamma_3$ & $\gamma_2$ & $\gamma_4$ & sample & method \\
   \hline
  \multirow{2}{*}{ AAA}   & 0.832                 & 2.569               & 0.361                  & -0.032            & 0.013            &                         &                      & theory\textsuperscript{\emph{a}} \\
                                        & \textbf{1.04}               & \textbf{3.2}        & \textbf{0.18}        &               &                       &                        &  on Platinum  & Our Work\\
\hline
  \multirow{8}{*}{ ABA}  & 0.84                  & 2.598               & 0.364                   & 0.319            & -0.014            & -0.177               &                      & theory\cite{hopping1991}\\
                                       &                           &                          &0.371                     &                      &                         &                           &exfoliation on SiO$_2$ & transport\textsuperscript{\emph{b}}\\
                                       & 1                        &  3.1                   &0.37                      &0.3                  &  -0.032             &0.04                    &exfoliation on SiO$_2$ & transport\textsuperscript{\emph{c}}\\
                                       & 0.99                  & 3.05                   &0.39                   &0.2                &                         &                           &on 6H-SiC(0001) & \multirow{2}{*}{ARPES\textsuperscript{\emph{d}}}\\
                                       & 1.13                   & 3.5                    &0.37                     &  0.2                &                         &                           &on 3C-SiC(111) &\\
                                       &1.06                     &3.27                  &0.44                     &                      &                         &                           &on 6H-SiC(0001) & ARPES\textsuperscript{\emph{e}}\\
                                        &\textbf{1.02}           &\textbf{3.15}          &\textbf{0.39}               &\textbf{0.25}          &                         &                           &on Platinum & Our Work\\
\hline
  \multirow{8}{*}{ ABC} &            &                         &0.502                      & -0.377             & -0.0171        &  -0.099            &                      & theory \textsuperscript{\emph{f}}\\
                                      &&&0.377&&&&exfoliation on SiO$_2$&transport\textsuperscript{\emph{b}}\\
                                      & 1                           & 3.09               &0.5                         &                         &                      &                          &exfoliation on SiO$_2$  & transport\cite{Zaliznyak2011}\\
                                      & 0.84                     &2.58                &0.34                      &   0.17              &                      &  0.04              &     & theory\textsuperscript{\emph{g}}\\
                                      & 0.93                      & 2.86               &0.38                      & 0.24                &                      &                          &on 6H-SiC(0001)  &\multirow{2}{*}{ARPES\textsuperscript{\emph{d}}}\\
                                      & 1.05                      & 3.24               &0.39                      & 0.24                 &                      &                          &on 3C-SiC(111)  & \\
                                      & 1                           & 3.08               &0.39                        &                         &                      &                          &14, 15 layers    & transport\cite{Faugeras2016}\\
                                      & \textbf{1.00}              & \textbf{3.1}        &\textbf{0.4}               &  \textbf{0.2}           &                      &                          &on Platinum  & Our Work\\
  \hline
  \hline
\end{tabular}
\textsuperscript{\emph{a}} Reference \citenum{Gonze1992}; \textsuperscript{\emph{b}} Reference \citenum{gap2011}; \textsuperscript{\emph{c}} Reference \citenum{hoppingExp2015}; \textsuperscript{\emph{d}} Reference \citenum{ARPES2013}; \textsuperscript{\emph{e}} Reference \citenum{bilayerOhta2006}; \textsuperscript{\emph{f}} Reference \citenum{TheoryBandABC2010}; \textsuperscript{\emph{g}} Reference \citenum{flatband2013}.
\end{table*}

To confirm the stacking sequences of  trilayer graphene and to reveal the interlayer and intralayer coupling of graphene in different stacking sequences, we use tight-binding model \cite{TheoryBand2008PRB} to fit the ARPES spectra (see the supplementary material for details). As shown in Fig. \ref{fig1}(a)-(c), \emph{Ab initio} numerical calculations \cite{TheoryBandABC2010,Gonze1992} and quantum capacitance measurements \cite{hoppingExp2015} suggest that hopping terms $\gamma_0$, $\gamma_1$ and $\gamma_3$ are enough to describe Bernal and rhombohedral stacked trilayer graphenes and  $\gamma_0$, $\gamma_1$ are enough for simple hexagonal stacking, so we only take these hopping terms into account respectively, where $\gamma _0$ is the nearest neighbor hopping in monolayer graphene, $\gamma_1$ is the vertical nearest interlayer hopping term, and $\gamma_3$ is the next nearest neighbor interlayer hopping term. The comparison of the experimental results and the fitting dispersions in Fig. \ref{fig3}(g)-(j) shows a good agreement with the dispersions for monolayer, simple hexagonal, Bernal and rhombohedral trilayer graphene respectively, confirming unambiguously the existence of all three trilayer stacking sequences. Besides, these hopping parameters reveal the interlayer and intralayer coupling of graphene and determine their band structure directly. The hopping integrals obtained from fitting the experimental data are listed in Table \ref{fit}.  It is well known that Fermi velocity is directly proportional to the nearest intralayer coupling ($\gamma_0$). The Fermi velocity obtained from $\gamma_0$ is ~$1.0\times10^6m/s$ for all stackings and is very close to that of pristine graphene \cite{Yao2015}.  The splitting between the bands increases with the vertical nearest interlayer coupling ($\gamma_1$) getting stronger.  Meanwhile, the next nearest neighbor interlayer coupling ($\gamma_3$) will tilt the bands for Bernal and rhombohedral stackings. For Bernal and rhombohedral stacking, the interlayer hopping integrals ($\gamma_1$, $\gamma_3$) agree well with previous theoretical \cite{flatband2013,Gonze1992} and experimental results including mechanically exfoliated trilayer graphene on SiO$_2$ \cite{ARPES2013,gap2011,hoppingExp2015} and synthesized trilayer graphene on 6H-SiC(0001) and  3C-SiC(111) substrates \cite{ARPES2007, ARPES2013} as shown in Table \ref{fit}. For simple hexagonal stacking (AAA) trilayer graphene, the interlayer hopping parameter ($\gamma_1$) is much weaker compared with Bernal or rhombohedral stacking. As far as we know, there are very few experimental reports about simple hexagonal stacking graphene \cite{Lee2008,Liu2009,Chang2008} and no hopping integral of AAA stacking has been reported before to compare with experimentally.

In summary, we have distinguished rhombohedral and Bernal stacked trilayer graphene spatially by their different band structures directly for the first time. Besides, we also observed simple hexagonal stacked trilayer graphene.  We have observed that the undoped graphene trilayer films are characterized by Dirac point locked at the Fermi level with different nesting depending on the stacking.   Clear dispersions of AAA, ABA and ABC stacking are obtained, and the experimental hopping parameters $\gamma_0$, $\gamma_1$ and $\gamma_3$ are obtained by fitting NanoARPES spectra.

\begin{acknowledgement}
We acknowledge support from the National Natural Science Foundation of China (Grant No. 11334006 and 11427903), from Ministry of Science and Technology of China(Grant No. 2015CB921001 and 2016YFA0301004) and Tsinghua University Initiative Scientific Research Program (2012Z02285). The Synchrotron SOLEIL is supported by the Centre National de la Recherche Scientifique (CNRS) and the Commissariat \` a l'Energie Atomique et aux Energies Alternatives (CEA), France.
\end{acknowledgement}

\section{METHODS}

The epitaxial graphene sample was grown by annealing the Pt(111) substrate in ultrahigh vacuum at elevated temperatures up to 1600 $^\circ$C using electron beam bombardment, which has been reported in our previous works \cite{Yao2015}. The sample was annealed at 450 $^\circ$C to clean the surface before the ARPES measurements. NanoARPES experiments were performed at the ANTARES beamline of the SOLEIL synchrotron, France. All ARPES data were taken at a photon energy of 100 eV with Scienta R4000 analyzer, using linearly polarized light. The vacuum was better than 2$\times10^{-10}$ Torr and the sample was kept at 80 K during the measurement.




\bibliography{reference}

\providecommand{\latin}[1]{#1}
\makeatletter
\providecommand{\doi}
  {\begingroup\let\do\@makeother\dospecials
  \catcode`\{=1 \catcode`\}=2\doi@aux}
\providecommand{\doi@aux}[1]{\endgroup\texttt{#1}}
\makeatother
\providecommand*\mcitethebibliography{\thebibliography}
\csname @ifundefined\endcsname{endmcitethebibliography}
  {\let\endmcitethebibliography\endthebibliography}{}
\begin{mcitethebibliography}{44}
\providecommand*\natexlab[1]{#1}
\providecommand*\mciteSetBstSublistMode[1]{}
\providecommand*\mciteSetBstMaxWidthForm[2]{}
\providecommand*\mciteBstWouldAddEndPuncttrue
  {\def\EndOfBibitem{\unskip.}}
\providecommand*\mciteBstWouldAddEndPunctfalse
  {\let\EndOfBibitem\relax}
\providecommand*\mciteSetBstMidEndSepPunct[3]{}
\providecommand*\mciteSetBstSublistLabelBeginEnd[3]{}
\providecommand*\EndOfBibitem{}
\mciteSetBstSublistMode{f}
\mciteSetBstMaxWidthForm{subitem}{(\alph{mcitesubitemcount})}
\mciteSetBstSublistLabelBeginEnd
  {\mcitemaxwidthsubitemform\space}
  {\relax}
  {\relax}

\bibitem[{Geim} and {Novoselov}(2007){Geim}, and {Novoselov}]{Graphene2007}
{Geim},~A.~K.; {Novoselov},~K.~S. \emph{Nat. Mater.} \textbf{2007}, \emph{6},
  183--191\relax
\mciteBstWouldAddEndPuncttrue
\mciteSetBstMidEndSepPunct{\mcitedefaultmidpunct}
{\mcitedefaultendpunct}{\mcitedefaultseppunct}\relax
\EndOfBibitem
\bibitem[{Aoki} and {Amawashi}(2007){Aoki}, and {Amawashi}]{TheoryBand2007}
{Aoki},~M.; {Amawashi},~H. \emph{Solid State Commun.} \textbf{2007},
  \emph{142}, 123--127\relax
\mciteBstWouldAddEndPuncttrue
\mciteSetBstMidEndSepPunct{\mcitedefaultmidpunct}
{\mcitedefaultendpunct}{\mcitedefaultseppunct}\relax
\EndOfBibitem
\bibitem[Oostinga \latin{et~al.}(2008)Oostinga, Heersche, Liu, Morpurgo, and
  Vandersypen]{bilayerOostinga2008}
Oostinga,~J.~B.; Heersche,~H.~B.; Liu,~X.; Morpurgo,~A.~F.; Vandersypen,~L.
  M.~K. \emph{Nat. Mater.} \textbf{2008}, \emph{7}, 151--157\relax
\mciteBstWouldAddEndPuncttrue
\mciteSetBstMidEndSepPunct{\mcitedefaultmidpunct}
{\mcitedefaultendpunct}{\mcitedefaultseppunct}\relax
\EndOfBibitem
\bibitem[Ohta \latin{et~al.}(2006)Ohta, Bostwick, Seyller, Horn, and
  Rotenberg]{bilayerOhta2006}
Ohta,~T.; Bostwick,~A.; Seyller,~T.; Horn,~K.; Rotenberg,~E. \emph{Science}
  \textbf{2006}, \emph{313}, 951--954\relax
\mciteBstWouldAddEndPuncttrue
\mciteSetBstMidEndSepPunct{\mcitedefaultmidpunct}
{\mcitedefaultendpunct}{\mcitedefaultseppunct}\relax
\EndOfBibitem
\bibitem[{Lui} \latin{et~al.}(2011){Lui}, {Li}, {Chen}, {Klimov}, {Brus}, and
  {Heinz}]{Lui2011Imaging}
{Lui},~C.~H.; {Li},~Z.; {Chen},~Z.; {Klimov},~P.~V.; {Brus},~L.~E.;
  {Heinz},~T.~F. \emph{Nano Lett.} \textbf{2011}, \emph{11}, 164--169\relax
\mciteBstWouldAddEndPuncttrue
\mciteSetBstMidEndSepPunct{\mcitedefaultmidpunct}
{\mcitedefaultendpunct}{\mcitedefaultseppunct}\relax
\EndOfBibitem
\bibitem[Zhang \latin{et~al.}(2016)Zhang, Han, Qiao, Tan, Wang, Zhang, and
  Tan]{Raman1}
Zhang,~X.; Han,~W.-P.; Qiao,~X.-F.; Tan,~Q.-H.; Wang,~Y.-F.; Zhang,~J.;
  Tan,~P.-H. \emph{Carbon} \textbf{2016}, \emph{99}, 118--122\relax
\mciteBstWouldAddEndPuncttrue
\mciteSetBstMidEndSepPunct{\mcitedefaultmidpunct}
{\mcitedefaultendpunct}{\mcitedefaultseppunct}\relax
\EndOfBibitem
\bibitem[Lui \latin{et~al.}(2013)Lui, Cappelluti, Li, and Heinz]{Raman2}
Lui,~C.~H.; Cappelluti,~E.; Li,~Z.; Heinz,~T.~F. \emph{Phys. Rev. Lett.}
  \textbf{2013}, \emph{110}, 185504\relax
\mciteBstWouldAddEndPuncttrue
\mciteSetBstMidEndSepPunct{\mcitedefaultmidpunct}
{\mcitedefaultendpunct}{\mcitedefaultseppunct}\relax
\EndOfBibitem
\bibitem[Lui \latin{et~al.}(2012)Lui, Malard, Kim, Lantz, Laverge, Saito, and
  Heinz]{Raman3}
Lui,~C.~H.; Malard,~L.~M.; Kim,~S.; Lantz,~G.; Laverge,~F.~E.; Saito,~R.;
  Heinz,~T.~F. \emph{Nano Lett.} \textbf{2012}, \emph{12}, 5539--44\relax
\mciteBstWouldAddEndPuncttrue
\mciteSetBstMidEndSepPunct{\mcitedefaultmidpunct}
{\mcitedefaultendpunct}{\mcitedefaultseppunct}\relax
\EndOfBibitem
\bibitem[Lui \latin{et~al.}(2015)Lui, Ye, Keiser, Barros, and He]{Raman4}
Lui,~C.~H.; Ye,~Z.; Keiser,~C.; Barros,~E.~B.; He,~R. \emph{Appl. Phys. Lett.}
  \textbf{2015}, \emph{106}, 041904\relax
\mciteBstWouldAddEndPuncttrue
\mciteSetBstMidEndSepPunct{\mcitedefaultmidpunct}
{\mcitedefaultendpunct}{\mcitedefaultseppunct}\relax
\EndOfBibitem
\bibitem[Lu \latin{et~al.}(2007)Lu, Chang, Huang, Ho, Hwang, and Lin]{Lu2007}
Lu,~C.~L.; Chang,~C.~P.; Huang,~Y.~C.; Ho,~J.~H.; Hwang,~C.~C.; Lin,~M.~F.
  \emph{J. Phys. Soc. Jpn.} \textbf{2007}, \emph{76}, 024701\relax
\mciteBstWouldAddEndPuncttrue
\mciteSetBstMidEndSepPunct{\mcitedefaultmidpunct}
{\mcitedefaultendpunct}{\mcitedefaultseppunct}\relax
\EndOfBibitem
\bibitem[Chang \latin{et~al.}(2008)Chang, Wang, Lu, Huang, Lin, and
  Chen]{Chang2008}
Chang,~C.~P.; Wang,~J.; Lu,~C.~L.; Huang,~Y.~C.; Lin,~M.~F.; Chen,~R.~B.
  \emph{J. Appl. Phys.} \textbf{2008}, \emph{103}, 103109\relax
\mciteBstWouldAddEndPuncttrue
\mciteSetBstMidEndSepPunct{\mcitedefaultmidpunct}
{\mcitedefaultendpunct}{\mcitedefaultseppunct}\relax
\EndOfBibitem
\bibitem[Gr\"uneis \latin{et~al.}(2008)Gr\"uneis, Attaccalite, Wirtz, Shiozawa,
  Saito, Pichler, and Rubio]{TheoryBand2008PRB}
Gr\"uneis,~A.; Attaccalite,~C.; Wirtz,~L.; Shiozawa,~H.; Saito,~R.;
  Pichler,~T.; Rubio,~A. \emph{Phys. Rev. B} \textbf{2008}, \emph{78},
  205425\relax
\mciteBstWouldAddEndPuncttrue
\mciteSetBstMidEndSepPunct{\mcitedefaultmidpunct}
{\mcitedefaultendpunct}{\mcitedefaultseppunct}\relax
\EndOfBibitem
\bibitem[Latil and Henrard(2006)Latil, and Henrard]{Latil2006}
Latil,~S.; Henrard,~L. \emph{Phys. Rev. Lett.} \textbf{2006}, \emph{97},
  036803\relax
\mciteBstWouldAddEndPuncttrue
\mciteSetBstMidEndSepPunct{\mcitedefaultmidpunct}
{\mcitedefaultendpunct}{\mcitedefaultseppunct}\relax
\EndOfBibitem
\bibitem[Guinea \latin{et~al.}(2007)Guinea, {Castro Neto}, and
  Peres]{Guinea2007}
Guinea,~F.; {Castro Neto},~A.~H.; Peres,~N. M.~R. \emph{Solid State Commun.}
  \textbf{2007}, \emph{143}, 116--122\relax
\mciteBstWouldAddEndPuncttrue
\mciteSetBstMidEndSepPunct{\mcitedefaultmidpunct}
{\mcitedefaultendpunct}{\mcitedefaultseppunct}\relax
\EndOfBibitem
\bibitem[Min and Macdonald(2008)Min, and Macdonald]{TheoryBand2008}
Min,~H.; Macdonald,~A.~H. \emph{Prog. Theor. Phys. Suppl.} \textbf{2008},
  \emph{176}, 227--252\relax
\mciteBstWouldAddEndPuncttrue
\mciteSetBstMidEndSepPunct{\mcitedefaultmidpunct}
{\mcitedefaultendpunct}{\mcitedefaultseppunct}\relax
\EndOfBibitem
\bibitem[Koshino(2010)]{TheoryBand2010}
Koshino,~M. \emph{Phys. Rev. B} \textbf{2010}, \emph{81}, 125304\relax
\mciteBstWouldAddEndPuncttrue
\mciteSetBstMidEndSepPunct{\mcitedefaultmidpunct}
{\mcitedefaultendpunct}{\mcitedefaultseppunct}\relax
\EndOfBibitem
\bibitem[Craciun \latin{et~al.}(2009)Craciun, Russo, Yamamoto, Oostinga,
  Morpurgo, and Tarucha]{Craciun2009}
Craciun,~M.~F.; Russo,~S.; Yamamoto,~M.; Oostinga,~J.~B.; Morpurgo,~a.~F.;
  Tarucha,~S. \emph{Nat. Nanotechnol.} \textbf{2009}, \emph{4}, 383--388\relax
\mciteBstWouldAddEndPuncttrue
\mciteSetBstMidEndSepPunct{\mcitedefaultmidpunct}
{\mcitedefaultendpunct}{\mcitedefaultseppunct}\relax
\EndOfBibitem
\bibitem[Koshino and McCann(2009)Koshino, and McCann]{koshino2009gate}
Koshino,~M.; McCann,~E. \emph{Phys. Rev. B} \textbf{2009}, \emph{79},
  125443\relax
\mciteBstWouldAddEndPuncttrue
\mciteSetBstMidEndSepPunct{\mcitedefaultmidpunct}
{\mcitedefaultendpunct}{\mcitedefaultseppunct}\relax
\EndOfBibitem
\bibitem[Koshino and McCann(2009)Koshino, and McCann]{TheoryBandABC2009}
Koshino,~M.; McCann,~E. \emph{Phys. Rev. B} \textbf{2009}, \emph{80},
  165409\relax
\mciteBstWouldAddEndPuncttrue
\mciteSetBstMidEndSepPunct{\mcitedefaultmidpunct}
{\mcitedefaultendpunct}{\mcitedefaultseppunct}\relax
\EndOfBibitem
\bibitem[Zhang \latin{et~al.}(2010)Zhang, Sahu, Min, and
  MacDonald]{TheoryBandABC2010}
Zhang,~F.; Sahu,~B.; Min,~H.; MacDonald,~A.~H. \emph{Phys. Rev. B}
  \textbf{2010}, \emph{82}, 035409\relax
\mciteBstWouldAddEndPuncttrue
\mciteSetBstMidEndSepPunct{\mcitedefaultmidpunct}
{\mcitedefaultendpunct}{\mcitedefaultseppunct}\relax
\EndOfBibitem
\bibitem[Avetisyan \latin{et~al.}(2009)Avetisyan, Partoens, and
  Peeters]{Avetisyan2009}
Avetisyan,~A.~A.; Partoens,~B.; Peeters,~F.~M. \emph{Phys. Rev. B}
  \textbf{2009}, \emph{80}, 195401\relax
\mciteBstWouldAddEndPuncttrue
\mciteSetBstMidEndSepPunct{\mcitedefaultmidpunct}
{\mcitedefaultendpunct}{\mcitedefaultseppunct}\relax
\EndOfBibitem
\bibitem[Avetisyan \latin{et~al.}(2010)Avetisyan, Partoens, and
  Peeters]{gap2010}
Avetisyan,~A.~A.; Partoens,~B.; Peeters,~F.~M. \emph{Phys. Rev. B}
  \textbf{2010}, \emph{81}, 115432\relax
\mciteBstWouldAddEndPuncttrue
\mciteSetBstMidEndSepPunct{\mcitedefaultmidpunct}
{\mcitedefaultendpunct}{\mcitedefaultseppunct}\relax
\EndOfBibitem
\bibitem[Lui \latin{et~al.}(2011)Lui, Li, Mak, Cappelluti, and Heinz]{gap2011}
Lui,~C.~H.; Li,~Z.; Mak,~K.~F.; Cappelluti,~E.; Heinz,~T.~F. \emph{Nat. Phys.}
  \textbf{2011}, \emph{7}, 944--947\relax
\mciteBstWouldAddEndPuncttrue
\mciteSetBstMidEndSepPunct{\mcitedefaultmidpunct}
{\mcitedefaultendpunct}{\mcitedefaultseppunct}\relax
\EndOfBibitem
\bibitem[Bao \latin{et~al.}(2011)Bao, Jing, {Velasco Jr}, Lee, Liu, Tran,
  Standley, Aykol, Cronin, Smirnov, Koshino, McCann, Bockrath, and
  Lau]{Bao2011}
Bao,~W.; Jing,~L.; {Velasco Jr},~J.; Lee,~Y.; Liu,~G.; Tran,~D.; Standley,~B.;
  Aykol,~M.; Cronin,~S.~B.; Smirnov,~D. \latin{et~al.}  \emph{Nat. Phys.}
  \textbf{2011}, \emph{7}, 948--952\relax
\mciteBstWouldAddEndPuncttrue
\mciteSetBstMidEndSepPunct{\mcitedefaultmidpunct}
{\mcitedefaultendpunct}{\mcitedefaultseppunct}\relax
\EndOfBibitem
\bibitem[Kopnin \latin{et~al.}(2013)Kopnin, Ij\"as, Harju, and
  Heikkil\"a]{flatband2013}
Kopnin,~N.~B.; Ij\"as,~M.; Harju,~A.; Heikkil\"a,~T.~T. \emph{Phys. Rev. B}
  \textbf{2013}, \emph{87}, 140503\relax
\mciteBstWouldAddEndPuncttrue
\mciteSetBstMidEndSepPunct{\mcitedefaultmidpunct}
{\mcitedefaultendpunct}{\mcitedefaultseppunct}\relax
\EndOfBibitem
\bibitem[Kopnin \latin{et~al.}(2011)Kopnin, Heikkil\"a, and
  Volovik]{flatband2011}
Kopnin,~N.~B.; Heikkil\"a,~T.~T.; Volovik,~G.~E. \emph{Phys. Rev. B}
  \textbf{2011}, \emph{83}, 220503\relax
\mciteBstWouldAddEndPuncttrue
\mciteSetBstMidEndSepPunct{\mcitedefaultmidpunct}
{\mcitedefaultendpunct}{\mcitedefaultseppunct}\relax
\EndOfBibitem
\bibitem[Pierucci \latin{et~al.}(2015)Pierucci, Sediri, Hajlaoui, Girard,
  Brumme, Calandra, Velezfort, Patriarche, Silly, Ferro, Souli\`{e}re,
  Marangolo, Sirotti, Mauri, and Ouerghi]{flatband2015}
Pierucci,~D.; Sediri,~H.; Hajlaoui,~M.; Girard,~J.-C.; Brumme,~T.;
  Calandra,~M.; Velezfort,~E.; Patriarche,~G.; Silly,~M.~G.; Ferro,~G.
  \latin{et~al.}  \emph{ACS Nano} \textbf{2015}, \emph{9}, 5432--5439\relax
\mciteBstWouldAddEndPuncttrue
\mciteSetBstMidEndSepPunct{\mcitedefaultmidpunct}
{\mcitedefaultendpunct}{\mcitedefaultseppunct}\relax
\EndOfBibitem
\bibitem[Scheike \latin{et~al.}(2012)Scheike, B{\"{o}}hlmann, Esquinazi,
  Barzola-Quiquia, Ballestar, and Setzer]{superScheike2012}
Scheike,~T.; B{\"{o}}hlmann,~W.; Esquinazi,~P.; Barzola-Quiquia,~J.;
  Ballestar,~A.; Setzer,~A. \emph{Adv. Mater.} \textbf{2012}, \emph{24},
  5826--5831\relax
\mciteBstWouldAddEndPuncttrue
\mciteSetBstMidEndSepPunct{\mcitedefaultmidpunct}
{\mcitedefaultendpunct}{\mcitedefaultseppunct}\relax
\EndOfBibitem
\bibitem[Otani \latin{et~al.}(2010)Otani, Takagi, Koshino, and
  Okada]{Otani2010a}
Otani,~M.; Takagi,~Y.; Koshino,~M.; Okada,~S. \emph{Appl. Phys. Lett.}
  \textbf{2010}, \emph{96}, 242504\relax
\mciteBstWouldAddEndPuncttrue
\mciteSetBstMidEndSepPunct{\mcitedefaultmidpunct}
{\mcitedefaultendpunct}{\mcitedefaultseppunct}\relax
\EndOfBibitem
\bibitem[Otani \latin{et~al.}(2010)Otani, Koshino, Takagi, and
  Okada]{Otani2010}
Otani,~M.; Koshino,~M.; Takagi,~Y.; Okada,~S. \emph{Phys. Rev. B}
  \textbf{2010}, \emph{81}, 161403\relax
\mciteBstWouldAddEndPuncttrue
\mciteSetBstMidEndSepPunct{\mcitedefaultmidpunct}
{\mcitedefaultendpunct}{\mcitedefaultseppunct}\relax
\EndOfBibitem
\bibitem[Coletti \latin{et~al.}(2013)Coletti, Forti, Principi, Emtsev,
  Zakharov, Daniels, Daas, Chandrashekhar, Ouisse, Chaussende, MacDonald,
  Polini, and Starke]{ARPES2013}
Coletti,~C.; Forti,~S.; Principi,~A.; Emtsev,~K.~V.; Zakharov,~A.~A.;
  Daniels,~K.~M.; Daas,~B.~K.; Chandrashekhar,~M. V.~S.; Ouisse,~T.;
  Chaussende,~D. \latin{et~al.}  \emph{Phys. Rev. B} \textbf{2013}, \emph{88},
  155439\relax
\mciteBstWouldAddEndPuncttrue
\mciteSetBstMidEndSepPunct{\mcitedefaultmidpunct}
{\mcitedefaultendpunct}{\mcitedefaultseppunct}\relax
\EndOfBibitem
\bibitem[Ohta \latin{et~al.}(2007)Ohta, Bostwick, McChesney, Seyller, Horn, and
  Rotenberg]{ARPES2007}
Ohta,~T.; Bostwick,~A.; McChesney,~J.~L.; Seyller,~T.; Horn,~K.; Rotenberg,~E.
  \emph{Phys. Rev. Lett.} \textbf{2007}, \emph{98}, 206802\relax
\mciteBstWouldAddEndPuncttrue
\mciteSetBstMidEndSepPunct{\mcitedefaultmidpunct}
{\mcitedefaultendpunct}{\mcitedefaultseppunct}\relax
\EndOfBibitem
\bibitem[Yao \latin{et~al.}(2015)Yao, Wang, Deng, Yang, Wu, Fedorov, Mo,
  Schwier, Zheng, Kojima, Iwasawa, Shimada, Jiang, Yu, Li, and Zhou]{Yao2015}
Yao,~W.; Wang,~E.; Deng,~K.; Yang,~S.; Wu,~W.; Fedorov,~A.~V.; Mo,~S.~K.;
  Schwier,~E.~F.; Zheng,~M.; Kojima,~Y. \latin{et~al.}  \emph{Phys. Rev. B}
  \textbf{2015}, \emph{92}, 115421\relax
\mciteBstWouldAddEndPuncttrue
\mciteSetBstMidEndSepPunct{\mcitedefaultmidpunct}
{\mcitedefaultendpunct}{\mcitedefaultseppunct}\relax
\EndOfBibitem
\bibitem[Sutter \latin{et~al.}(2009)Sutter, Sadowski, and Sutter]{Sutter}
Sutter,~P.; Sadowski,~J.~T.; Sutter,~E. \emph{Phys. Rev. B} \textbf{2009},
  \emph{80}, 245411\relax
\mciteBstWouldAddEndPuncttrue
\mciteSetBstMidEndSepPunct{\mcitedefaultmidpunct}
{\mcitedefaultendpunct}{\mcitedefaultseppunct}\relax
\EndOfBibitem
\bibitem[Gao \latin{et~al.}(2011)Gao, Pan, Huang, Hu, Zhang, Guo, Du, and
  Gao]{GaoHJ}
Gao,~M.; Pan,~Y.; Huang,~L.; Hu,~H.; Zhang,~L.~Z.; Guo,~H.~M.; Du,~S.~X.;
  Gao,~H.-J. \emph{Appl. Phys. Lett.} \textbf{2011}, \emph{98}, 033101\relax
\mciteBstWouldAddEndPuncttrue
\mciteSetBstMidEndSepPunct{\mcitedefaultmidpunct}
{\mcitedefaultendpunct}{\mcitedefaultseppunct}\relax
\EndOfBibitem
\bibitem[Damascelli \latin{et~al.}(2003)Damascelli, Hussain, and
  Shen]{ARPESReview}
Damascelli,~A.; Hussain,~Z.; Shen,~Z.-X. \emph{Rev. Mod. Phys.} \textbf{2003},
  \emph{75}, 473--541\relax
\mciteBstWouldAddEndPuncttrue
\mciteSetBstMidEndSepPunct{\mcitedefaultmidpunct}
{\mcitedefaultendpunct}{\mcitedefaultseppunct}\relax
\EndOfBibitem
\bibitem[Charlier \latin{et~al.}(1991)Charlier, Michenaud, Gonze, and
  Vigneron]{hopping1991}
Charlier,~J.~C.; Michenaud,~J.~P.; Gonze,~X.; Vigneron,~J.~P. \emph{Phys. Rev.
  B} \textbf{1991}, \emph{44}, 13237--13249\relax
\mciteBstWouldAddEndPuncttrue
\mciteSetBstMidEndSepPunct{\mcitedefaultmidpunct}
{\mcitedefaultendpunct}{\mcitedefaultseppunct}\relax
\EndOfBibitem
\bibitem[Zhang \latin{et~al.}(2011)Zhang, Zhang, Camacho, Khodas, and
  Zaliznyak]{Zaliznyak2011}
Zhang,~L.; Zhang,~Y.; Camacho,~J.; Khodas,~M.; Zaliznyak,~I. \emph{Nat. Phys.}
  \textbf{2011}, \emph{7}, 953--957\relax
\mciteBstWouldAddEndPuncttrue
\mciteSetBstMidEndSepPunct{\mcitedefaultmidpunct}
{\mcitedefaultendpunct}{\mcitedefaultseppunct}\relax
\EndOfBibitem
\bibitem[Henni \latin{et~al.}(2016)Henni, Collado, Nogajewski, Molas, Usaj,
  Balseiro, Orlita, Potemski, and Faugeras]{Faugeras2016}
Henni,~Y.; Collado,~H.; Nogajewski,~K.; Molas,~M.; Usaj,~G.; Balseiro,~C.;
  Orlita,~M.; Potemski,~M.; Faugeras,~C. \emph{arXiv preprint arXiv:1603.03611}
  \textbf{2016}, \relax
\mciteBstWouldAddEndPunctfalse
\mciteSetBstMidEndSepPunct{\mcitedefaultmidpunct}
{}{\mcitedefaultseppunct}\relax
\EndOfBibitem
\bibitem[Charlier \latin{et~al.}(1992)Charlier, Michenaud, and
  Gonze]{Gonze1992}
Charlier,~J.-C.; Michenaud,~J.-P.; Gonze,~X. \emph{Phys. Rev. B} \textbf{1992},
  \emph{46}, 4531--4539\relax
\mciteBstWouldAddEndPuncttrue
\mciteSetBstMidEndSepPunct{\mcitedefaultmidpunct}
{\mcitedefaultendpunct}{\mcitedefaultseppunct}\relax
\EndOfBibitem
\bibitem[Wu \latin{et~al.}(2015)Wu, Han, Lin, Zhu, He, Xu, Chen, Lu, Ye, Han,
  Wu, Long, Shen, Huang, Wang, He, Cai, Lortz, Su, and Wang]{hoppingExp2015}
Wu,~Z.; Han,~Y.; Lin,~J.; Zhu,~W.; He,~M.; Xu,~S.; Chen,~X.; Lu,~H.; Ye,~W.;
  Han,~T. \latin{et~al.}  \emph{Phys. Rev. B} \textbf{2015}, \emph{92},
  075408\relax
\mciteBstWouldAddEndPuncttrue
\mciteSetBstMidEndSepPunct{\mcitedefaultmidpunct}
{\mcitedefaultendpunct}{\mcitedefaultseppunct}\relax
\EndOfBibitem
\bibitem[Lee \latin{et~al.}(2008)Lee, Lee, Ahn, Kim, Wilson, and John]{Lee2008}
Lee,~J.-K.; Lee,~S.-C.; Ahn,~J.-P.; Kim,~S.-C.; Wilson,~J. I.~B.; John,~P.
  \emph{J. Chem. Phys.} \textbf{2008}, \emph{129}, 234709\relax
\mciteBstWouldAddEndPuncttrue
\mciteSetBstMidEndSepPunct{\mcitedefaultmidpunct}
{\mcitedefaultendpunct}{\mcitedefaultseppunct}\relax
\EndOfBibitem
\bibitem[Liu \latin{et~al.}(2009)Liu, Suenaga, Harris, and Iijima]{Liu2009}
Liu,~Z.; Suenaga,~K.; Harris,~P. J.~F.; Iijima,~S. \emph{Phys. Rev. Lett.}
  \textbf{2009}, \emph{102}, 015501\relax
\mciteBstWouldAddEndPuncttrue
\mciteSetBstMidEndSepPunct{\mcitedefaultmidpunct}
{\mcitedefaultendpunct}{\mcitedefaultseppunct}\relax
\EndOfBibitem
\end{mcitethebibliography}
\end{document}